# Mimivirus Gene Promoters Exhibit an Unprecedented Conservation among all Eukaryotes

Karsten Suhre[1*], Stéphane Audic[1*], Jean-Michel Claverie[1,2]

[1] Information Génomique & Structurale, CNRS UPR2589, IBSM, 13402 Marseille;
[2]University of Méditerranée, School of Medicine, 13385, Marseille, France

[*] These two authors contributed equally to this work.

Running head : Promoters are Uniquely Conversed in Mimivirus

**ABSTRACT**

The initial analysis of the recently sequenced genome of *Acanthamoeba polyphaga Mimivirus*, the largest known double-stranded DNA virus, predicted a proteome of size and complexity more akin to small parasitic bacteria than to other nucleo-cytoplasmic large DNA viruses, and identified numerous functions never before described in a virus. It has been proposed that the Mimivirus lineage could have emerged before the individualization of cellular organisms from the 3 domains of life. An exhaustive *in silico* analysis of the non-coding moiety of all known viral genomes, now uncovers the unprecedented perfect conservation of a AAAATTGA motif in close to 50% of the Mimivirus genes. This motif preferentially occurs in genes transcribed from the predicted leading strand and is associated with functions required early in the viral infectious cycle, such as transcription and protein translation. A comparison with the known promoter of unicellular eukaryotes, in particular amoebal protists, strongly suggests that the AAAATTGA motif is the structural equivalent of the TATA box core promoter element. This element is specific to the Mimivirus lineage, and may correspond to an ancestral promoter structure predating the radiation of the eukaryotic kingdoms. This unprecedented conservation of core promoter regions is another exceptional features of Mimivirus, that again raises the question of its evolutionary origin.

## INTRODUCTION

The recent discovery and genome analysis of A*canthamoeba polyphaga Mimivirus*, the largest known DNA virus, challenged much of the accepted dogma regarding viruses (1, 2). Its particle size (>400 nm), genome length (1.2 million bp) and huge gene repertoire (911 protein coding genes) all contribute to blur the established boundaries between viruses and the smallest parasitic cellular organisms. Phylogenetic analysis placed the evolutionary origin of Mimivirus prior to the emergence of the extant eukaryotic kingdoms, raising the possibility that large DNA viruses might define a domain distinct from the three other domains of life, Eucarya, Archea and Bacteria (2). The exceptionally large gene content of the Mimivirus genome, that includes key protein translation genes as well as a very diverse set of enzymes belonging to different metabolic pathways, is consistent with the hypothesis that Mimivirus (and other large DNA viruses) evolved from a free living organism through a genome reduction process akin to the one experienced by parasitic intracellular bacteria (3-5). However, less radical scenarios have been proposed, such as the recruitment of numerous host acquired genes complementing a set of core genes common to all Nucleo-cytoplasmic large DNA viruses (NCLDVs) (6). In the latter case, the acquired open reading frames would require to be transferred with its own promoter, or be put under the control of a suitable viral promoter to function properly. These questions about the origin of the Mimivirus genes prompted us to systematically analyze the DNA sequences immediately 5'upstream of each of Mimivirus ORFs in search for transcriptional motifs. Unexpectedly, this analysis revealed the presence of the perfectly conserved octamer AAAATTGA in the putative core promoter regions of near half of all Mimivirus genes. The same analysis, applied to the genomes of other large DNA viruses, confirmed that such an homogeneity of the core promoter sequences is a unique characteristics of Mimivirus.

## RESULTS & DISCUSSION

### A highly conserved motif in the 5'upstream regions of Mimivirus genes

The annotation of the Mimivirus genome sequence defines 911 predicted protein coding genes, plus an additional set of 347 less convincing unidentified reading frames (URFs).

Intergenic regions have an average size of 157 nt with a standard deviation of 113 nt (excluding the two-tailed 5% most extreme data points). Starting from the predicted gene map, we searched for conserved motifs within the 150-nucleotide regions upstream of each of the predicted translation start codon (ATG) of the 911 genes. These sub-sequences were identified (i) as statistically over-represented short oligomers, or (ii) using the Gibbs sampler algorithm implemented in MEME (7).

In contrast to the common difficulty of extracting well conserved signals from eukaryotic promoters (8), our results were unexpectedly clear cut: we found that 403 of the 911 (45%) predicted Mimivirus genes exhibited a strictly conserved AAAATTGA motif within their 150-nucleotide upstream region. The statistical significance of this finding in the context of the exact nucleotide composition of Mimivirus genome was assessed as follows. We cut the Mimivirus genome into 15752 consecutive 150-nucleotide long segments (using both strands) and counted the occurrences of the above motif in each of them. Only 661 (4%) of these segments were found to contain the AAAATTGA motif at least once. Such a strong preferential occurrence of this motif in the 5'upstream region of Mimivirus genes is indeed highly significant (Fisher Exact $p < 10^{-194}$, for the table [(403, 911)(661, 15752)]. Overall, 60% of the occurrences of this motif are located in front of a gene. Moreover, the distribution of the motif within these 5' upstream regions is non random, with most of them being located in a narrow range, between 50 and 110 nucleotides from the translation start (Figure 1). A sequence logo (9) including the flanking regions of these AAAATTGA motifs shows that they are preceded by 2-3 AT-rich positions and followed by 8-9 AT-rich positions (Figure 2).

The same set of 150-nucleotide upstream regions was also analyzed using the more sophisticated Gibbs sampler approach, as implemented in MEME (7). A position-weight matrix (PWM) that corresponds to a motif very similar to the previous AAAATTGA was identified (MEME motif; Figure S1). Using a score cutoff of 1000 (based on the bi-modal distribution of the PWM scores; supplementary Figure S2), 446 genes (49%) with a motif were detected. For comparison, only 464 (3%) of the above 15752 consecutive genomic segments exhibit the MEME motif (score≥1000). Most of them are thus found upstream of a gene coding region. (Fisher Exact p-value $< 10^{-280}$). We note that no other predominant motif was found in the upstream regions of genes not exhibiting the AAAATTGA motif. Finally, we verified that the prevalence of the AAAATTGA motif in the 150-nucleotide segment upstream of the ORFs was not a mere statistical consequence of their high A+T content. When these sequences were randomized, the AAAATTGA octamer was only found 22.7 times (compared to 403) and the MEME motif 7.2 times (compared to 446) in average over 100 repetitions of the randomization protocol.

**The motif is less frequent in the more hypothetical URFs**

The same analysis was performed with the 347 unlikely genes, annotated URFs, in the Mimivirus genome. Only 20 (6%) of them exhibited the AAAATTGA motif in their 150-nucleotide upstream region. Moreover, many of them did not fall not within the [-110, -50] range, or were not flanked by AT rich positions. Consistently, only 11 (3%) of the URFs exhibited the previously defined MEME motif (score>1000). This motif distribution is not significantly different from the one observed in the 15752 consecutive 150-nucleotide long segments covering the whole genome.

**The presence/absence of the motif does not correlate with the intergenic distance**

We verified that the occurrences of the AAAATTGA motif were not trivially linked to the distance between successive ORFs. Figure 3 shows that the proportion of genes exhibiting the motif is fairly constant across the whole range of intergenic distances, except for the smallest ones (< 60 nt). It is worth to notice that a fraction of these short distances might be artifactual, corresponding to cases where the proximal ATG does not coincide with the actual translation start. The net result of this artifact is to slightly minimize the number of genes exhibiting the AAAATTGA motif.

**The motif significantly correlates with genes transcribed from the leading strand**

We then examined other possible correlations between the presence of the AAAATTGA motifs and the positions of the corresponding genes within the genome. Overall, the distribution of Mimivirus protein coding genes does not exhibit any significant strand bias: 450 genes are found on the positive strand (R-genes) and 461 on the negative strand (L-genes; Fisher Exact p-value = 0.8). The distribution of the AAAATTGA motif is similarly unbiased with 196 occurrence in front of the R-genes and 207 in front of the L-genes. The MEME motif exhibits a similar distribution with 217 in front of R-genes and 229 in front of L-genes (score>1000). There is thus no significant global strand preference for the occurrence of the upstream motif (Fisher Exact p-value = 0.8).

However, a previous analysis (2) identified a putative origin of replication (OR) of the chromosome near position 380,000 (between genes L294 and L295). On the basis of this prediction, one can distinguish a "leading" strand with 578 genes transcribed away from the OR, and a "lagging" strand from which 333 are transcribed. Interestingly, the AAAATTGA motif occurs significantly more frequently (Fisher Exact p-value = 0.027) in front of the genes transcribed from the leading strand (281/578 = 48.6%) than in front of the genes transcribed from the lagging strand (122/333 = 36.6%). A similar asymmetry is found for the MEME motif (313/578=54.2% and 133/313 = 39.9%; Fisher Exact p-value = 0.015).

We also examined whether the presence of AAAATTGA motif might correlate with certain categories of gene functions (Table 1). 157 Mimivirus genes can be associated to one of the COG (10) functional classes. We found that most translation- and transcription-related genes exhibited the AAAATTGA motif. This was also true of genes related to nucleotide transport and metabolism. In contrast, the motif was absent from the upstream region of most genes related to DNA replication, recombination, and repair as well as from genes classified in the cell envelope biogenesis / outer membrane COGs. Overall, the upstream motif does not occur more frequently in front of genes associated with a functional annotations (88/232 = 38%) compared to anonymous ones (315/679 = 46%) (Tables S1).

**NCLDV core genes and the AAAATTGA motif**

Iyer et al. (6) identified a set of homologues genes which have been identified in all or most members of the four main NCLDV groups: Iridoviridae, Asfarviridae, Phycodnaviriae and Poxviridae. Some of these "core" genes are also found in baculoviridae and phages. These core genes are divided in 4 classes, from the most to the least conserved. Remarkably, we found that none of the 9 class I core genes found in Mimivirus have the exact octamer motif in their 5' upstream region, and only 2 exhibit the MEME motif (Table S2). In contrast, all but one of the six class II core genes have the octamer motif and the MEME motif. The motif distribution for the class III (7 /11) and Class IV (10/16) core genes is more balanced. However, none of the above distributions significantly differ from 446/911 ratio observed for all Mimivirus genes (Fisher exact test p-value>0.1).

**The presence of such a highly conserved 5' upstream motif is unique to Mimivirus**

For many years, the 5' upstream region of eukaryotic genes has been under intense scrutiny in many different organisms from different kingdom (e.g. fungi, plant and metazoan) in an attempt to decipher the sequence-based signal involved in the initiation and the regulation of the transcription process(11, 12). The sole common result that emerged from these numerous studies is that sequence conservation is the exception rather than the rule in the 5'upstream regions of eukaryotic genes (13, 14). Other recent studies have confirmed this lack of conservation in several lineage of parasitic protists (15). However, large eukaryotic DNA viruses have been much less studied than their cellular counterparts. The analysis as applied to the Mimivirus genome was thus performed on the genomes of NCLDVs of the Iridiviridae, Phycodnaviridae, and Poxviridae families, to establish the pattern of sequence conservation in the 5'upstream regions of their genes. Our results show that none of these other NCLDVs exhibit a pattern of conservation comparable to the one observed for Mimivirus (Table S3). For instance, only 30 out of 178 (17%) genes of the *Invertebrate iridescent virus 6* (CIV, Iridoviridae) have a conserved AAAATTGA motif within their 150 nt upstream region. 14 out of 231 (6%) *Fowlpox* (FOP, Poxviridae) genes, and 10 of 218 (4.5%) *Amsacta moorei entomopoxvirus* (AME, Poxviridae) genes also exhibit this motif. In all other NCLDVs less or no occurrences of this motif were detected. Note however, that 47 of the 218 (22%) AME genes display a conserved TTTTGAAA motif (Table S3). Finally, an analysis by Gibbs sampling of all viral genomes containing more than 100 annotated genes showed that the Mimivirus pattern of 5'upstream motif conservation is truly unique among large DNA viruses. Such a conservation is also absent from archea viruses such as AFV1 (16). In absence of experimental data we cannot formally exclude that the intergenic AAAATTGA motif may act on both, upstream and downstream adjacent genes. However, a symmetrical analysis of the 150-nucleotide long intergenic regions 3' downstream of each gene identified only half (203 of 403) of the previously identified motifs, and exhibited no preferential location for these motifs with respect to the preceding Stop codon (supplementary Fig. S3). These results are thus in favor of a 5' polarity of function.

**The conserved AAAATTGA motif is likely to be a main core promoter element**

Analyses of the structure and expression of a number of genes from amoebal protists have shown that they are expressed in single transcription units, and that few of them have introns (15). As the mechanisms of gene expression used by a virus and its host must be compatible, it is reasonable to propose that the short genome regions (157 nt average) separating two consecutive Mimivirus ORFs contain most of the promoter sequence information. The eukaryotic core promoter includes DNA elements that can extend 35 bp upstream and/or downstream of the transcription initiation site. Most core promoter elements appear to interact directly with components of the basal transcription machinery. In metazoan, the most conserved and recognizable core promoter element is the TATA box, located 25 to 30 bp of the transcription start site, and present in about one third of the human genes (17). The average position (-60 from the predicted initiator ATG) of the conserved octamer found in Mimivirus is consistent with it playing a role similar to the TATA box for the expression of the viral genes, provided the 3'untranslated regions (5'UTR) of viral mRNA are about 30 nucleotide-long in average. Such a compact promoter/3'UTR structure has been observed in amitochondriate protists such as *Giardia intestinalis* or *Entamoeba histolytica* (15). The sequences of the TATA box-like element of these protozoa are also different from the 5'-TATATAAG-3' consensus identified in the other eukaryotic kingdoms. Accordingly we propose that the AAAATTGA motif found in 50% of Mimivirus intergenic regions is the virus TATA box-like motif.

**Mimivirus TATA box-like motif is not prevalent in amoebal organisms**

Interestingly, the Mimivirus TATA box-like motif does not bear a particular resemblance (if any) with the different TATA box-like consensus sequences (if any) that have been identified in various protozoan. For instance, *E. histolytica* TATA box-like consensus is TATTTAAA (15). Using the available data from a genomic survey sequencing of *Acanthamoeba castellanii* (Anderson, I.J. and Loftus, B.J., Gene discovery in the *Acanthamoeba castellanii* genome, unpublished; ~19Mb sequence data available at http://www.ncbi.nlm.nih.gov/), we verified that the AAAATTGA motif is not particularly prevalent in the genome of the closest (partially sequenced) relative of the Mimivirus host *Acanthamoeba polyphaga*. In addition, no significant difference in codon usage between the sets of genes harboring or not harboring the AAAATTGA motif could be detected. Symmetrically, of the 29 Mimivirus genes most likely to have been acquired by lateral transfer (18), 15 exhibits the AAAATTGA motif and 14 do not have it. These results argue against the hypothesis that this promoter element, together with the large proportion of associated Mimivirus genes, were simply acquired from its host. Interestingly clusters of paralogous genes that have been produced by multiple round of duplications from an ancestral Mimivirus gene conserved the AAAATTGA motifs amidst the divergence of their respective promoter regions. The example of the large gene cluster L175-L185 is shown in Figure 4.

**The two types of Mimivirus promoters might correspond to early *vs.* late functions**

Homologues of the 3 main proteins involved in the formation of the transcription pre-initiation complex have been identified in Mimivirus genome: the two RNA polymerase II subunits (Rbp1 & Rpb2) and the TFIID (TATA-box binding) initiation factor. The corresponding amino acid sequences are extremely distant from their closest matches: *Candida albicans* Rpb1 (34% identical), *Dictyostelium discoideum* Rpb2 (36% identical) and *Plasmodium falciparum* TFIID (24% identical). We propose that Mimivirus unique TATA box-like sequence AAAATTGA might have co-evolved and be recognized by a pre-initiation complex including these divergent Mimivirus proteins. A second type of promoter – highly degenerate – might then be recognized by the host pre-initiation complex, or involving a combination of Mimivirus and host encoded transcription factors. This hypothesis is consistent with the preferential occurrence of the AAAATTGA (type one promoter) in front of Mimivirus genes encoding functions required for the early (or late-early) phase of viral infection (transcription, nucleotide transport, and protein translation) (Table 1). According to this scenario, the corresponding genes could be transcribed in the host cytoplasm. Conversely, the AAAATTGA motif is preferentially absent from the promoter of Mimivirus genes encoding “late” functions such as DNA replication and particle biogenesis and assembly (Table 1). These genes could be transcribed in the host nucleus.

## CONCLUSION

Our bioinformatics and comparative genomics study revealed a unique feature of Mimivirus among the eukaryotic domain: the presence of a highly conserved AAAATTGA motif in the immediate 5’ upstream region of 50% of its protein encoding genes. By analogy with the known promoter structures of unicellular eukaryotes, in particular amoebal organisms, we propose that this motif corresponds to a TATA box-like core promoter element. This element, and its conservation, appear to be specific of the Mimivirus lineage, and might correspond to an ancestral promoter structure predating the radiation of the eukaryotic kingdoms (2). Mimivirus genes exhibiting this type of promoter might be ancestral as well. Interestingly, this is the case of all translation apparatus- related genes for the first time identified in a virus (four amino-acyl tRNA synthetases, mRNA cap binding protein, translation factor eF-Tu, and tRNA methyltransferase), as well as Mimivirus Topoisomerase 1A (bacterial type). However, it is also possible, but less likely, that horizontally acquired ORFs could have been inserted downstream of a preexisting AAAATTGA motif.

More genomic data on amoebal species, other protozoa, and their associated viruses is now needed to better understand the evolutionary scenario through which the unique promoter structure of Mimivirus genes might have emerged. Our findings can now be used to guide the experimental characterization of Mimivirus transcription units, and identify the transcription factors involved in the recognition of its uniquely conserved promoter element.

## ACKNOWLEDGEMENTS

This work was supported by CNRS and a grant from the French National Genopole Network.

## TABLE AND FIGURE LEGENDS

**Table 1 :** Number of 'COGed' Mimivirus genes with or without conserved AAAATTGA motif (only COG classes that contain at least 3 Mimivirus genes are shown). Numbers corresponding to genes with or without a MEME motif are given in brackets.

**Figure 1 :** The distribution of the position of the AAAATTGA motif with respect to the predicted gene start shows the location of the conserved element around 50-110 nt..

**Figure 2 :** The sequence logo (based on 400 genes with a strictly conserved AAAATTGA motif ; 3 genes with a motif that is less than 20 nt from the predicted translation start were not included in the computation of this logo) shows the conserved AT-rich neighborhood of the exactly conserved AAAATTGA octamer (see Figure S1 for the corresponding logo of the MEME motif).

**Figure 3 :** No relation between the presence/absence of the AAAATTGA motif and the distance between two genes can be identified in the size distribution of Mimivirus intergenic regions (except for the virtual absence of the motif in intergenic regions that are too short in order to host it).

**Figure 4 :** The alignment of the intergenic region of the paralogous gene cluster L175-L185 shows the perfect conservation of the AAAATTGA motif within intergenic regions that have otherwise more diverged. Note that the AAAATTGA motif is part of the C-terminal of gene L185 (indicated by white X's).

## SUPPLEMENTARY TABLE AND FIGURE LEGENDS

**Table S1 :** List of functionally annotated Mimivirus genes. Genes with an AAAATTGA and a MEME motif in the 150nt upstream region are marked by 'X'; genes that display only a AAAATTGA motif by 'A'; genes with only a MEME motif by 'M'.

**Table S2 :** List of Mimivirus NCLDV core gene set members with and without the AAAATTGA motif; annotation of the presence or absence of a motif is as in Table S1.

**Table S3:** AAAATTGA-related conserved 8-mers in other large DNA viruses.

**Figure S1:** MEME sequence logo for the 446 genes having a PWM score > 1000.

**Figure S2 :** Distribution of the PWM scores (496 out of 911 genes have a score > 0; 446 have a score >1000). The bi-modality of the distribution supports the choice of our score cutoff at 1000.

**Figure S3 :** The distribution of the position of the AAAATTGA motif within the 150-nucleotide long intergenic segment 3' downstream of each gene. The position profile exhibits no conserved location with respect to the 5' upstream gene stop codon (see Fig. 1 for comparison).

**Table 1**

| with motif | without motif | Total | COG function |
|---|---|---|---|
| 7 (9) | 23 (21) | 30 | DNA replication, recombination, and repair |
| 0 (0) | 9 (9) | 9 | Cell envelope biogenesis, outer membrane |
| 3 (3) | 5 (5) | 8 | Amino acid transport and metabolism |
| 5 (5) | 11 (11) | 16 | Posttranslational modification, protein turnover, chaperones |
| 1 (1) | 2 (2) | 3 | Lipid metabolism |
| 34 (38) | 26 (22) | 60 | Function unknown or general function prediction only |
| 2 (2) | 1 (1) | 3 | Secondary metabolites biosynthesis, transport, and catabolism |
| 5 (5) | 0 (0) | 5 | Nucleotide transport and metabolism |
| 7 (7) | 2 (2) | 9 | Transcription |
| 6 (5) | 1 (2) | 7 | Translation, ribosomal structure and biogenesis |

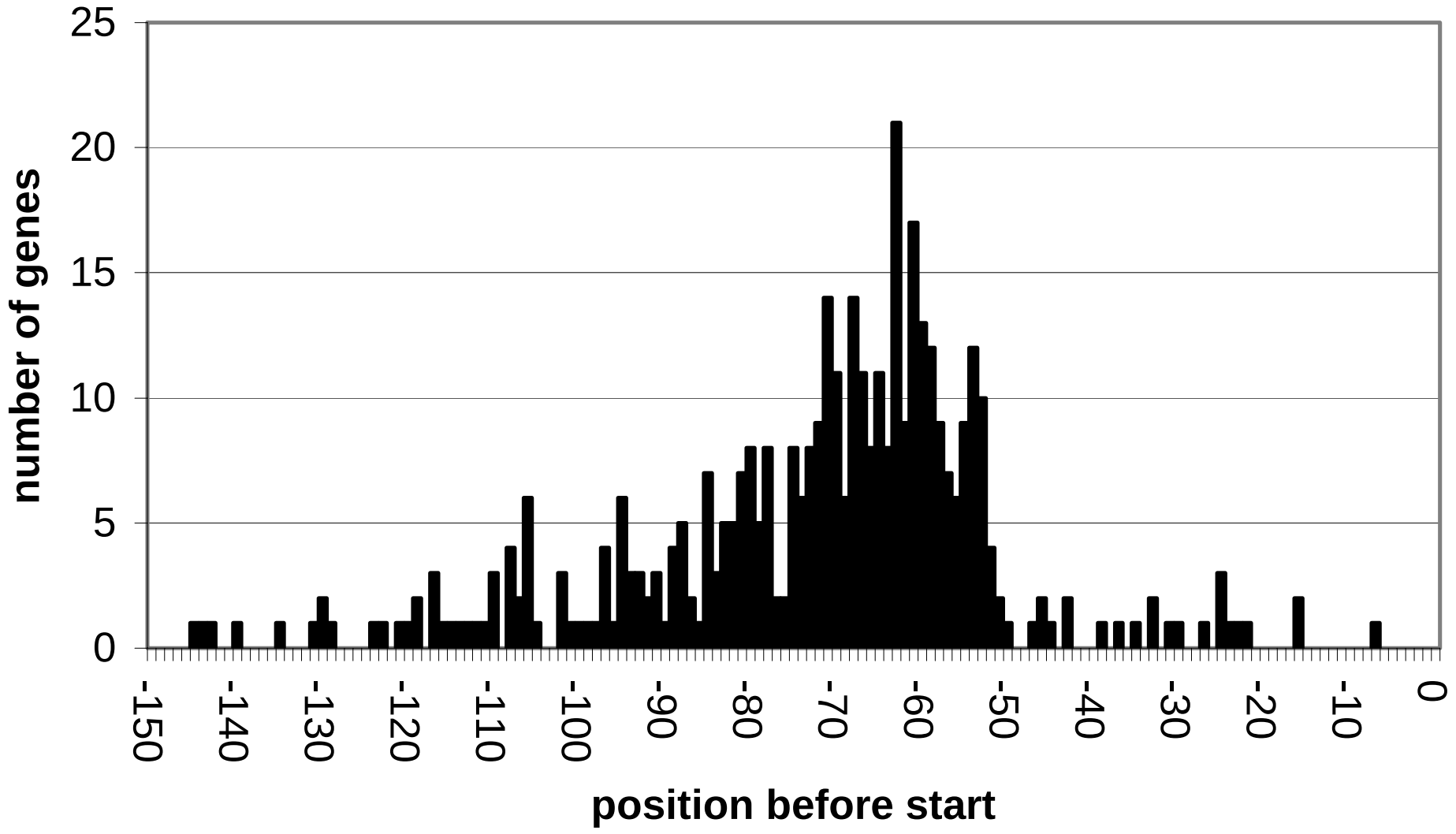

**Figure 1**

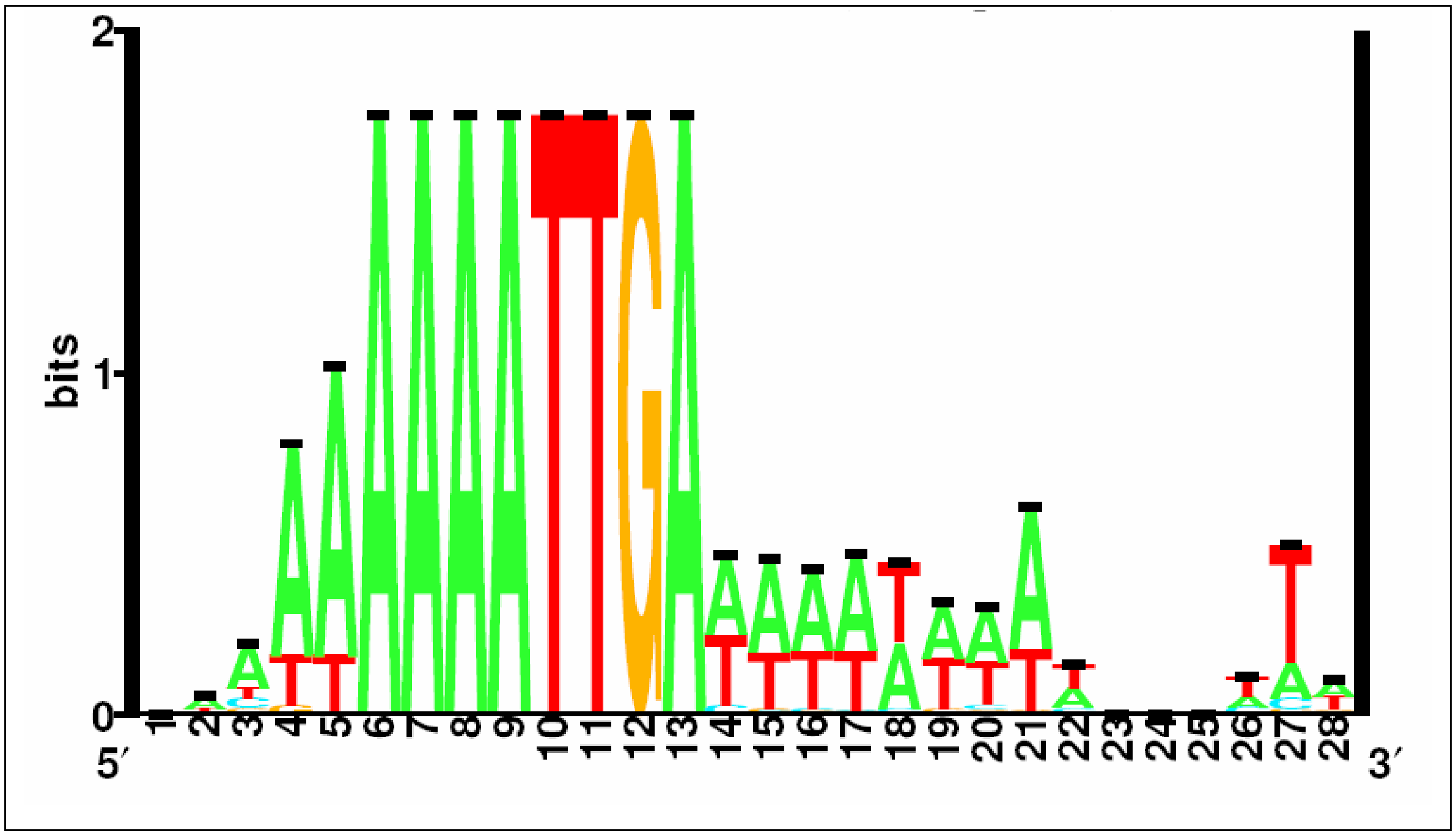
bits
2
1
0
5′
3′
1 2 3 4 5 6 7 8 9 10 11 12 13 14 15 16 17 18 19 20 21 22 23 24 25 26 27 28

**Figure 2**

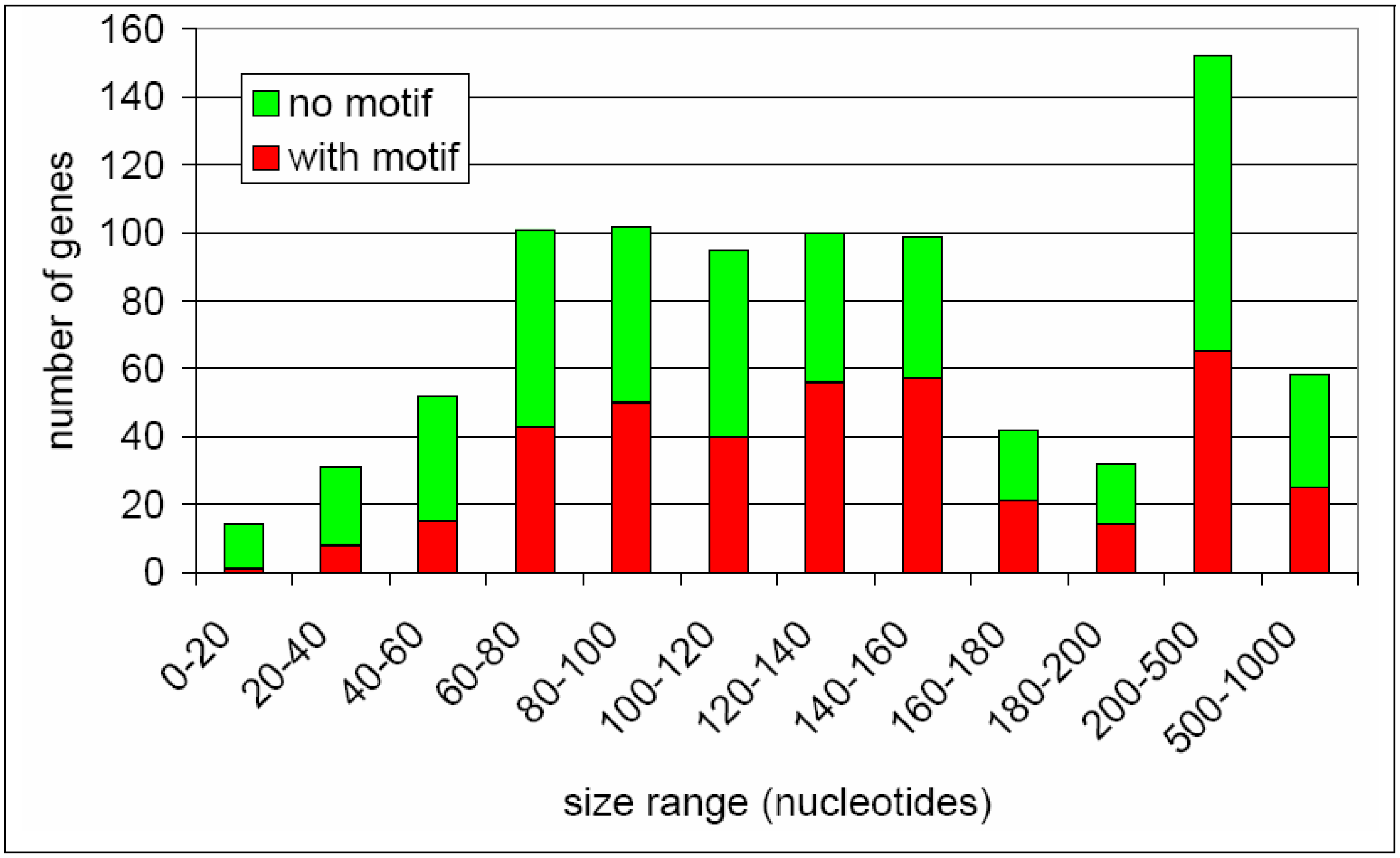

**Figure 3**

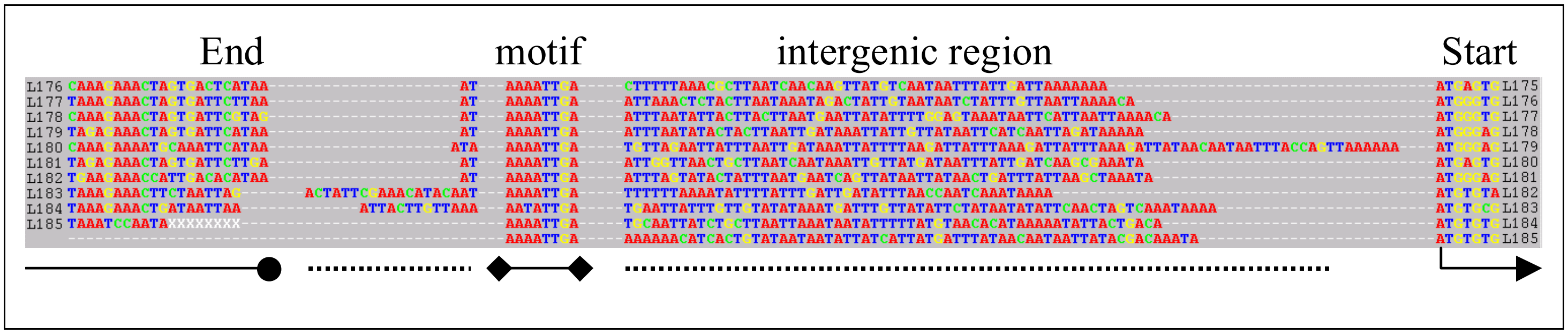

**Figure 4**

# Table S1

| Gene | Motif | Annotation |
|---|---|---|
| R1 | | Replication Origin binding protein |
| L2 | | Bro family N terminal domain containing protein |
| L4 | X | KilA N-terminal Domain, N1R P28 DNA binding |
| R8 | | Helicase similar to Origin binding protein |
| L30 | A | KilA N-terminal Domain, N1R P28 DNA binding |
| L32 | X | KilA N-terminal Domain, N1R P28 DNA binding |
| L33 | X | KilA N-terminal Domain, N1R P28 DNA binding |
| L37 | X | BroA, KilA-N term |
| R61 | X | WD-repeat family protein |
| L65 | | Virion-associated membrane protein |
| L71 | | Collagen triple helix repeat containing protein |
| L79 | | putative transposase |
| R80 | | putative resolvase |
| L84 | | WD-repeat family protein |
| L103 | | putative resolvase |
| R104 | | putative transposase |
| L108 | X | Proliferating cell nuclear antigen (PCNA) |
| L124 | X | Tyrosyl-tRNA synthetase |
| L128 | X | Choline dehydrogenase |
| R135 | | Choline dehydrogenase or related protein |
| L136 | | Sugar transaminase dTDP-4-amino-4,6-dideoxyglucose biosynthesis |
| L137 | | contains glycosyl-transferase domain (PFAM) |
| L138 | | Contain glycosyl-transferase domain (PFAM) |
| R139 | | contains glycosyl-transferase domain (PFAM) |
| R141 | | GDP mannose 4,6-dehydratase (hydroxysteorid dehydrogenase?) |
| L142 | | contains hexapeptidase domain (PFAM) |
| L153 | | similar to A273L from Paramecium bursaria Chlorella virus |
| L164 | X | Cysteinyl-tRNA synthetase |
| R186 | X | putative transposase |
| L193 | | putative glycosyl-transferase |
| R194 | | Topoisomerase I (pox-like) |
| R195 | | Glutaredoxin (ESV128 type) |
| R196 | | Collagen triple helix repeat containing protein |
| R197 | | Probable cytidine deaminase |
| L205 | X | Contains protein kinase domain (PFAM) |
| L206 | | Helicase III VV D5-type ATPase (C-term) |
| L207 | X | Helicase III VV D5-type ATPase (N-term) |
| R209 | M | probable RNA polymerase subunit 6 |
| R214 | | RAS family GTPase |
| L221 | X | Topoisomerase I (bacterial type) |
| L230 | | Procollagen-lysine,2-oxoglutarate 5-dioxygenase |
| L232 | M | Contains protein kinase domain (PFAM) |
| L233 | | putative Zn-dependent peptidase |
| L235 | A | RNA polymerase subunit 5 |
| R239 | | Collagen triple helix repeat containing protein |
| R240 | | Collagen triple helix repeat containing protein |
| R241 | | Collagen triple helix repeat containing protein |
| L244 | X | Rpb2 |
| L245 | | HNH endonuclease |
| L247 | X | HNH endonuclease |
| L249 | X | putative uracil-DNA glycosylase |
| L250 | X | putative transcription initiation factor IIB |
| L251 | | Lon domain protease |
| L254 | | Heat shock 70Kd HSP |
| L258 | X | Thymidine kinase |
| R259 | X | Contains DUF549 domain (PFAM) |
| R260 | X | Dnaj-like protein |
| R266 | | Dnaj-like protein |
| L268 | X | Contains protein kinase domain (PFAM) |
| L275 | | Contains K+ channel tetramerisation domain (PFAM) |
| L276 | X | Putative oxoglutarate malate carrier protein |
| R296 | | Endonuclease IV |
| R299 | | contain Ribonuclease H domain (PFAM) |
| L300 | | contains Endo excinuclease amino terminal domain (PFAM) |
| R301 | | Uncharacterized protein (Chilo iridescent virus 380R) |
| R303 | | NAD-dependent DNA ligase |
| R307 | | Contains Protein phosphatase 2C domain (PFAM) |
| R311 | X | BIR domain (Chilo iridescent virus 193R) |
| L312 | X | Ribonucleotide reductase small subunit |
| R313 | X | Ribonucleotide reductase large subunit |
| L315 | | Hydrolysis of DNA containing ring-opened N7 methylguanine |
| L316 | | Glucosamine 6-phosphate N-acetyltransferase |
| L318 | | DNA polymerase family X |
| R319 | X | putative ubiquitin-specific protease |
| R322 | M | DNA polymerase (B family) |
| L323 | | Myristoylated virion protein A |
| R325 | | Metal-dependent hydrolase (Chilo iridescent virus 136R) |
| R339 | X | TFII-like transcription factor |
| R343 | X | RNAse III (dsRNA binding motif) |
| L344 | X | WD-repeat family protein |
| R350 | | putative transcription termination factor, VV D6R helicase |
| R354 | | Lambda-type exonuclease |
| L359 | | DNA mismatch repair ATPase MutS |
| R362 | | contain Thioredoxin domain (PFAM) |
| R363 | | Contains glycosyl-transferase domain (PFAM) |
| L364 | X | SW1 SNF2 helicase (MSV224) |
| R366 | | Contains helicase domaine (PFAM) |
| R368 | | putative thiol oxidoreductase |
| L373 | | putative glycosyltransferase |
| L377 | | putative NTPase I |
| R380 | | Contains Leucine Rich Repeat domain (PFAM) |
| R382 | X | mRNA Capping Enzyme |
| L386 | X | putative endonuclease of the XPG family |
| L388 | X | contains Endonuclease Exonuclease phosphatase domain (PFAM) |
| L393 | | Bos Taurus HSP70 |
| R395 | X | similar to EsV-1-87 [Ectocarpus siliculosus virus] |
| L396 | M | VV A18 helicase |
| R398 | | Contains Calcineurin-like phosphoesterase domain (PFAM) |
| R400 | | S T protein kinase, similar to Paramecium bursaria chlorella virus 1 A617R |
| R405 | X | tRNA (Uracil-5-)-methyltransferase |
| R406 | X | Alkylated DNA repair |
| R407 | X | tRNA (uracil-5-)-methyltransferase |
| L410 | | Similar to poxvirus P4B major core protein |
| R418 | X | NDK synthesis of nucleoside triphosphates |
| R419 | | putative dehydrogenase |
| L425 | | Capsid protein D13L (4 paralogs) |
| R429 | | PBCV1-A494R-like, 9 paralogs |
| R431 | X | Contains 3'-5' exonuclease domain (PFAM) |
| L432 | | Predicted HD phosphohydrolase |
| R436 | X | Putative serine threonine protein kinase |
| L437 | | VV A32 virion packaging ATPase |
| R439 | | Capsid protein |
| R440 | | Capsid protein |
| R441 | | Capsid protein |
| R443 | | Contains thioredoxin domain (PFAM) |
| L444 | | Contains ADP-ribosylglycohydrolase domain (PFAM) |
| R445 | | Contains Dnaj domain (PFAM) |
| L446 | | Patatin-like phospholipase (463L) |
| R449 | | Uncharacterized protein (Paramecium bursaria |
| R450 | | A1L transcription factor |
| R453 | X | TATA-box binding protein (TBP) |
| R458 | X | putative translation initiation factor |
| L460 | X | ubiquitin-conjugating enzyme |
| R464 | | Translation initiation factor SUI1 |
| L469 | M | Polynucleotide phosphatase kinase |
| R470 | M | DNA directed RNA polymerase subunit L |
| R475 | X | Asparagine synthase (glutamine hydrolysing) |
| R476 | X | Contains ATPase domain (PFAM) |
| L477 | | Cathepsin B, Cystein protease |
| L478 | X | putative replication factor C |
| R480 | X | Topoisomerase II |
| R486 | | Contains 2 PAN domains (PFAM) |
| R493 | X | Proliferating Cell Nuclear Antigen |
| L496 | A | Translation initiation factor 4E, (mRNA cap binding) |
| R497 | X | Thymidylate synthase |
| L498 | | Zn-dependent alcohol dehydrogenase |
| L499 | X | putative replication factor C |
| R501 | X | Rpb1 |
| R510 | X | putative replication factor C subunit |
| L511 | A | Contains FtsJ-like methyltransferase domain (PFAM) |
| R512 | X | Deoxynucleotide monophosphate kinase |
| R517 | | Contains protein kinase domain (PFAM) |
| R518 | | Contains protein kinase domain (PFAM) |
| R519 | | putative peptidase, contains peptidase M13 domain (PFAM) |
| L524 | X | MuT-like NTP pyrophosphohydrolase |
| R526 | | putative triacylglycerol lipase |
| R530 | M | putative DNA helicase |
| L532 | X | Contains cytochrome p450 domain (PFAM) |
| L538 | | Helicase conserved C-terminal domain (PFAM) |
| L540 | | VVI8 helicase |
| R541 | X | K+ channel tetramerisation domain |
| L543 | X | ADP-ribosyltransferase (DraT) |
| L544 | | Transcription initiation factor TFIIB |
| R548 | | Contains thioredoxin domain (PFAM) |
| R555 | | putative DNA repair protein |
| R563 | | Contains helicase conserved C-terminal domain (PFAM) |
| R565 | | Glutamine synthetase (Glutamate-amonia ligase) |
| R568 | X | Contains UvrD REP helicase domain (PFAM) |
| R569 | | Contains exonuclease domain (PFAM) |
| R571 | | Patatin-like phospholipase (similar to Chilo iridescent virus 463L) |
| L572 | | Contains ATPase domain (PFAM) |
| R592 | X | Contains helicase conserved C-terminal domain (PFAM) |
| L593 | | prolyl 4-hydroxylase |
| R595 | | putative phospholipase carboxylesterase family protein |
| R596 | | Thiol oxidoreductase E10R |
| L605 | | peptidylprolyl isomerase |
| R610 | | proline rich protein |
| L612 | | Mannose-6P isomerase |
| R614 | X | Contains SPFH domain Band 7 family (PFAM) |
| L615 | | Contains FYVE zinc finger domain (PFAM) and |
| L619 | | putative glucosamine--fructose-6-phosphate aminotransferase |
| L620 | | Contains Patatin-like phospholipase domain (PFAM) |
| L621 | | putative N-myristoyltransferase |
| R622 | | Dual specificity S Y phosphatase |
| R624 | X | GTP binding elongation factor eF-Tu |
| L628 | | Cyt-b5, Cytochrome b5-like Heme Steroid binding domain |
| L630 | | putative Ubiquitin-conjugating enzyme E2 |
| R632 | | putative N-acylsphingosine amidohydrolase |
| R639 | X | Methyonyl-tRNA synthetase |
| R644 | | putative Phosphatidylethanolamine-binding protein (PFAM) |
| R654 | | Contains glycosyltransferase domain (PFAM) |
| R655 | | Contain Glycosyltransferase domain (PFAM) |
| R663 | X | Arginyl-tRNA synthetase |
| R665 | X | putative oxidoreductase |
| L668 | | Collagen triple helix repeat containing protein |
| L669 | | Collagen triple helix repeat containing protein |
| L670 | | Contains Protein kinase domain (PFAM) and Cyclin, N-terminal domain (PFAM) |
| L673 | X | Contains protein kinase domain (PFAM) and Cyclin, N-terminal domain (PFAM) |
| L687 | | Endonuclease for the repair of UV-irradiated DNA |
| R689 | | N-acetylglucosamine-1-phosphate uridyltransferase |
| R693 | X | Methylated-DNA-protein-cysteinemethyltransferase |
| R700 | | Contains Serpin (serine protease inhibitor) domain (PFAM) |
| R707 | | P13-like protein |
| R708 | | Fructose-2,6-bisphosphatase |
| L709 | X | Contains Ubiquitin-conjugating enzyme domain (PFAM) |
| L716 | | GMP synthase (Glutamine-hydrolyzing) |
| L720 | | Hydrolysis of DNA containing ring-opened N7 methylguanine |
| R721 | | Similar to CheD, Chemotaxis protein |
| R726 | | Peptide chain release factor eRF1 |
| R730 | | Contains ATPase domain (PFAM) |
| L733 | X | ABC transporter ATP-binding domain |
| R739 | X | WD-repeat family protein |
| R749 | X | Contains homeobox domain (PFAM) |
| R756 | X | similar to Predicted Fe-S-cluster redox enzyme (COG) |
| L770 | | putative transposase |
| R771 | | putative resolvase |
| L775 | | Contains 2 PAN domains (PFAM) |
| L780 | | dTDP-4-dehydrorhamnose reductase |
| L783 | A | WD-repeat family protein |
| R795 | X | Contains ubiquitin-conjugating enzyme domain (PFAM) |
| L805 | | MACRO domain (splicing related) |
| R807 | M | 7-dehydrocholesterol reductase |
| L808 | X | Lanosterol 14-alpha-demethylase |
| R818 | | Contains 2 protein kinase domains (PFAM) |
| L823 | X | Proliferating Cell Nuclear Antigen |
| R824 | X | Putative 5'(3')-deoxyribonucleotidase |
| R826 | X | Contains 2 protein kinase domains (PFAM) |
| R831 | X | Contains 2 protein kinase domains (PFAM) |
| R832 | X | putative alcohol dehydrogenase (N-term) |
| R833 | | putative alcohol dehydrogenase (C-term) |
| R854 | | putative transposase |
| R856 | X | Contains 6 TPR domains (PFAM) |
| R877 | | putative outer membrane lipoprotein |
| R878 | X | KilA N-terminal Domain, N1R P28 DNA binding |
| R879 | M | KilA N-terminal Domain, N1R P28 DNA binding |
| R892 | X | putative oxidoreductase (C-term) |
| L893 | | putative oxidoreductase (C-term) |
| L894 | | putative oxidoreductase (N-term) |
| R904 | X | KilA N-terminal Domain, N1R P28 DNA binding |
| L905 | | putative methyl-transferase |
| L906 | X | Acetylcholinesterase |

## Table S2

| Class | Gene | Motif | Annotation |
|---|---|---|---|
| I | L206 | | Helicase III VV D5-type ATPase (C-term) |
| I | L396 | M | VV A18 helicase |
| I | L425 | | Capsid protein D13L (4 paralogs) |
| I | L437 | | VV A32 virion packaging ATPase |
| I | R322 | M | DNA polymerase (B family) |
| I | R350 | | putative transcription termination factor, VV D6R helicase |
| I | R400 | | S T protein kinase, similar to Paramecium bursaria chlorella virus 1 A617R |
| I | R450 | | A1L transcription factor |
| I | R596 | | Thiol oxidoreductase E10R |
| II | L312 | X | Ribonucleotide reductase small subunit |
| II | L323 | | Myristoylated virion protein A |
| II | L524 | X | MuT-like NTP pyrophosphohydrolase |
| II | R313 | X | Ribonucleotide reductase large subunit |
| II | R339 | X | TFII-like transcription factor |
| II | R493 | X | Proliferating Cell Nuclear Antigen |
| III | L244 | X | Rpb2 |
| III | L364 | X | SW1 SNF2 helicase (MSV224) |
| III | L37 | X | BroA, KilA-N term |
| III | L65 | | Virion-associated membrane protein |
| III | R195 | | Glutaredoxin (ESV128 type) |
| III | R311 | X | BIR domain (Chilo iridescent virus 193R) |
| III | R382 | X | mRNA Capping Enzyme |
| III | R429 | | PBCV1-A494R-like, 9 paralogs |
| III | R480 | X | Topoisomerase II |
| III | R501 | X | Rpb1 |
| III | R622 | | Dual specificity S Y phosphatase |
| IV | L235 | A | RNA polymerase subunit 5 |
| IV | L258 | X | Thymidine kinase |
| IV | L32 | X | KilA N-terminal Domain, N1R P28 DNA binding |
| IV | L33 | X | KilA N-terminal Domain, N1R P28 DNA binding |
| IV | L37 | X | BroA, KilA-N term |
| IV | L4 | X | KilA N-terminal Domain, N1R P28 DNA binding |
| IV | L446 | | Patatin-like phospholipase (463L) |
| IV | L477 | | Cathepsin B, Cystein protease |
| IV | L540 | | VVI8 helicase |
| IV | L805 | | MACRO domain (splicing related) |
| IV | R141 | | GDP mannose 4,6-dehydratase (hydroxysteorid dehydrogenase?) |
| IV | R301 | | Uncharacterized protein (Chilo iridescent virus 380R) |
| IV | R303 | | NAD-dependent DNA ligase |
| IV | R325 | | Metal-dependent hydrolase (Chilo iridescent virus 136R) |
| IV | R343 | X | RNAse III (dsRNA binding motif) |
| IV | R354 | | Lambda-type exonuclease |
| IV | R449 | | Uncharacterized protein (Paramecium bursaria |
| IV | R497 | X | Thymidylate synthase |
| IV | R571 | | Patatin-like phospholipase (similar to Chilo iridescent virus 463L) |
| IV | R878 | X | KilA N-terminal Domain, N1R P28 DNA binding |
| IV | R879 | M | KilA N-terminal Domain, N1R P28 DNA binding |
| IV | R904 | X | KilA N-terminal Domain, N1R P28 DNA binding |

**Table S3**

| | *Acanthamoeba polyphaga Mimivirus* | *Invertebrate iridescent virus 6* | *Fowlpox virus* | *Amsacta moorei entomopox virus* |
|---|---|---|---|---|
| annotated genes | 911 | 178 | 231 | 218 |
| with AAAATTGA motif | 403 | 30 | 14 | 10 |
| chunks per strand (150 nt) | 7877 | 1417 | 1924 | 1550 |
| chunks with motif | 664 | 61 | 22 | 56 |
| AAAATTGA (best Mimi) | **403** | 30 | **14** | 10 |
| AAATTGAA (best CIV) | 206 | **46** | 12 | 14 |
| TTTTGAAA (best AME) | 28 | 18 | 3 | **47** |

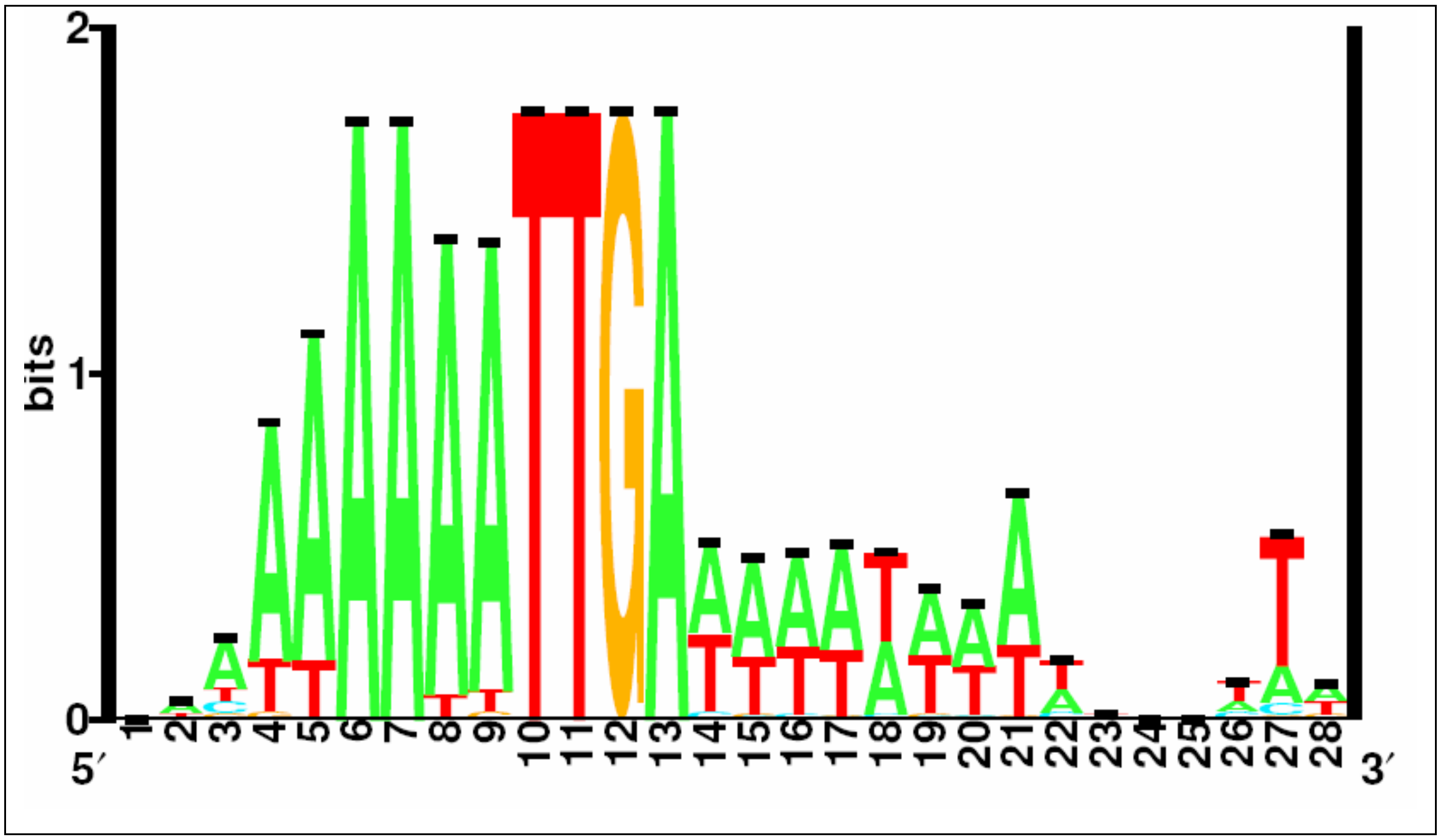
2
bits
1
0
5′
1 2 3 4 5 6 7 8 9 10 11 12 13 14 15 16 17 18 19 20 21 22 23 24 25 26 27 28
3′

**Figure S1**

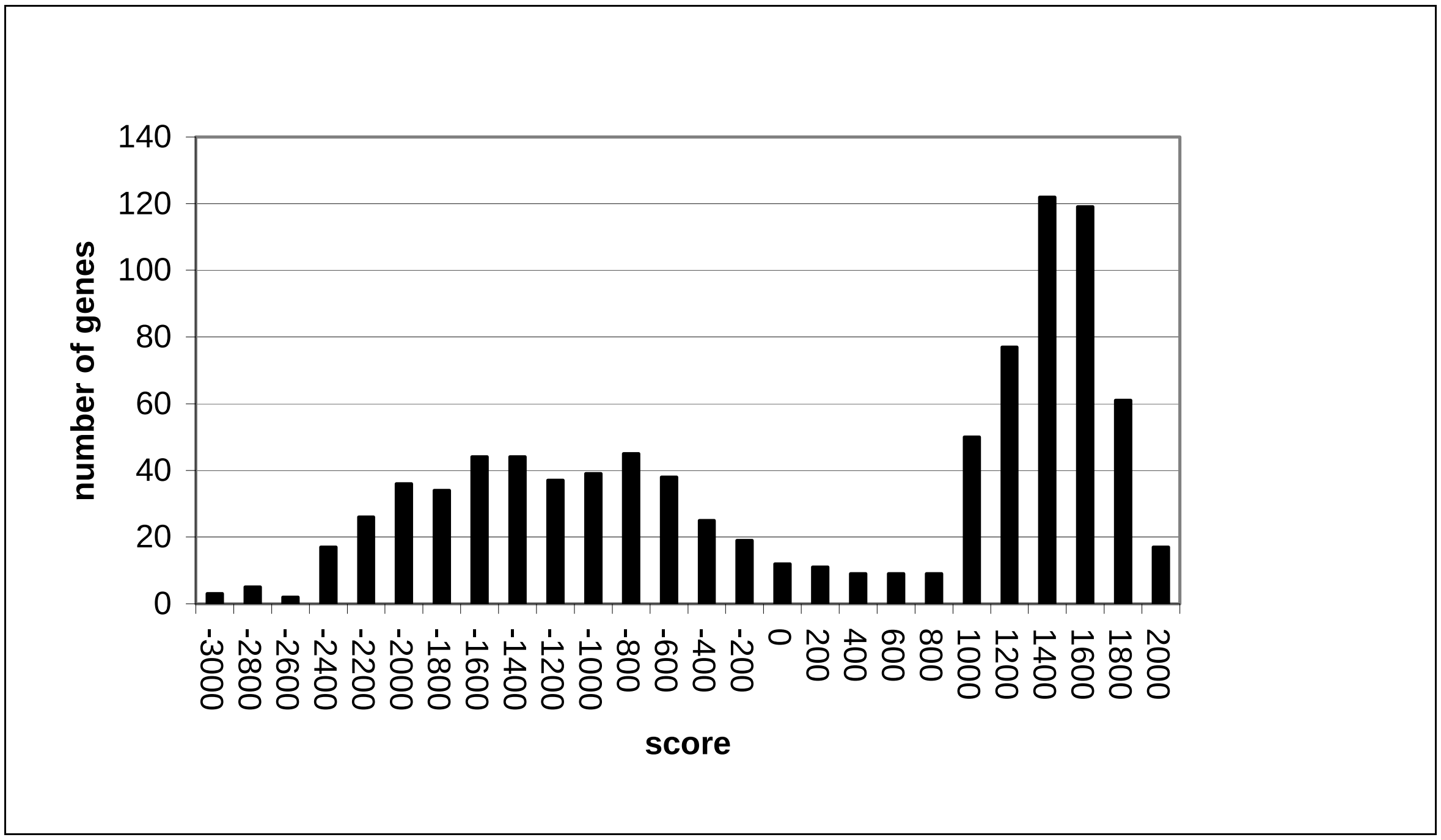

**Figure S2**

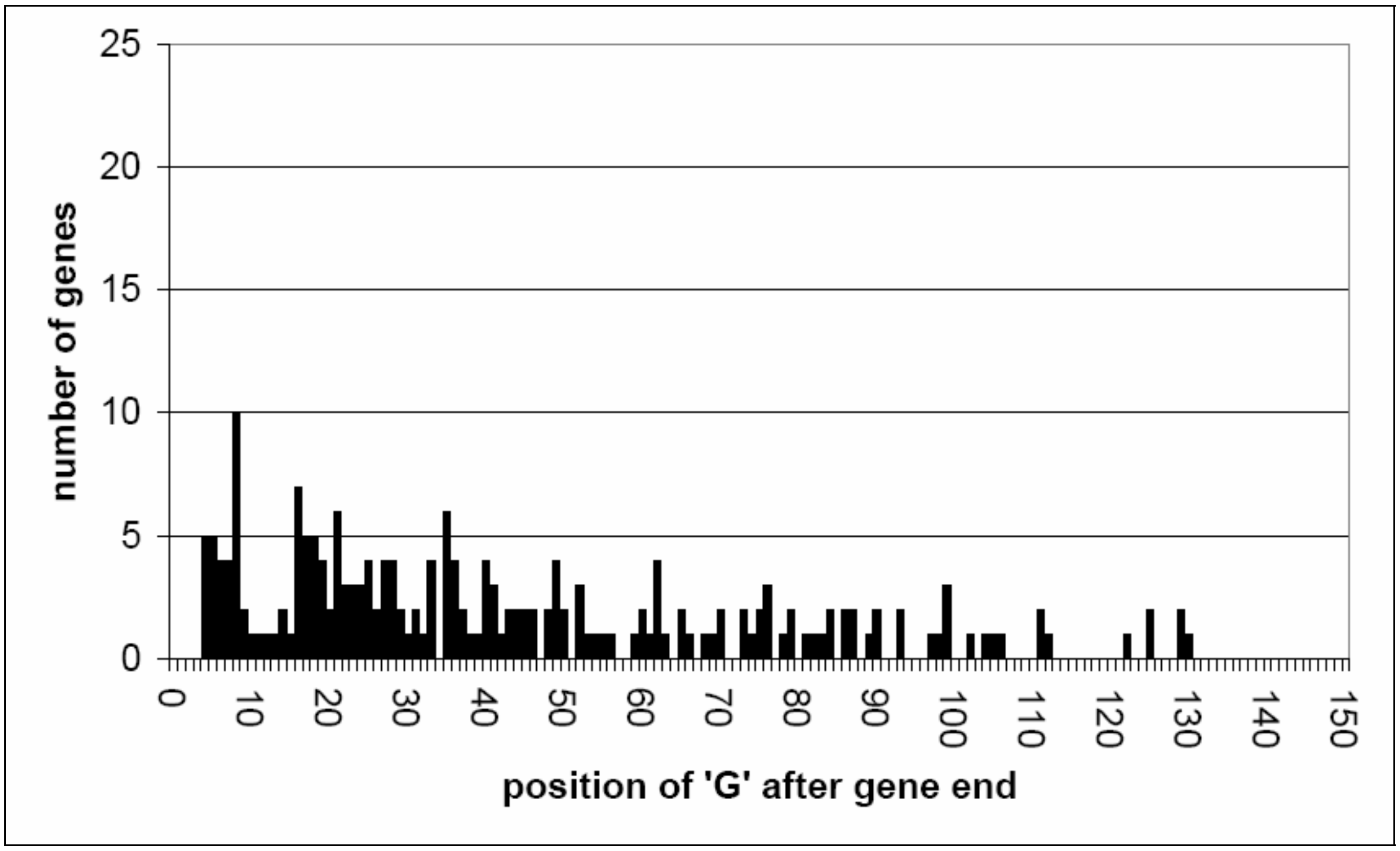

**Figure S3**